\documentclass[pdflatex,sn-mathphys-num]{sn-jnl}

\usepackage{graphicx}%
\usepackage{multirow}%
\usepackage{amsmath,amssymb,amsfonts}%
\usepackage{amsthm}%
\usepackage{mathrsfs}%
\usepackage[title]{appendix}%
\usepackage{xcolor}%
\usepackage{textcomp}%
\usepackage{manyfoot}%
\usepackage{booktabs}%
\usepackage{tabularx}
\usepackage{makecell}

\usepackage{algorithm}
\usepackage{algorithmic}
\usepackage{listings}%

\usepackage{tikz}
\usetikzlibrary{positioning, calc}

\usepackage{url} 

\usepackage{subcaption}

\theoremstyle{thmstyleone}%
\theoremstyle{thmstyletwo}%

\theoremstyle{thmstylethree}%

\begin{document}

\title{Bridging Semantics and Strategy: A Dual-Stream Graph Network for Equitable Negotiation Forecasting}


 \author{\fnm{Moirangthem Tiken} \sur{Singh}}\email{tiken.m@dibru.ac.in}

\affil{\orgdiv{Department of CSE},
\orgname{Dibrugarh University Institute of Engineering and Technology (DUIET), Dibrugarh University},
\orgaddress{\city{Dibrugarh}, \state{Assam}, \country{India}}}


\abstract{Forecasting outcomes in mixed-motive negotiations requires integrating explicit linguistic cues with latent strategic constraints, such as budgets and alternatives. Existing computational models often fail to adapt to varying task structures and may not adequately account for distributive considerations present in historical training data. This study proposes a unified framework to adaptively fuse semantic and strategic signals while incorporating reflective modeling of utility disparities. We introduce the Semantic-Temporal Graph Fusion Network (ST-GFN), a dual-stream architecture that processes textual dialogue with transformer encoders and economic states with Graph Attention Networks, connected via a dynamic gated fusion mechanism. Evaluated on contrasting benchmarks—the linguistically oriented DealOrNoDeal and the strategy-oriented CaSiNo, ST-GFN exhibits strong adaptability. The model dynamically adjusts modality weighting, emphasizing linguistic cues in free-form settings (z $\approx$ 0.97) and increasing reliance on strategic constraints in structured tasks (z $\approx$ 0.73). A fairness-regularized composite loss is incorporated to penalize deviations from ground-truth utility gaps. Results demonstrate a 43.8\% reduction in Inequality Discrepancy in high-disparity environments with minimal impact on accuracy, alongside improved performance in high-variance domains. These findings suggest that reflective regularization can enhance both predictive reliability and equitable representation in negotiation forecasting, supporting the design of transparent Group Decision and Negotiation Support Systems (GDNSS).}

\keywords{Automated Negotiation, Information Fusion, Graph Neural Networks, Strategic Reasoning, Algorithmic Fairness}



\maketitle

\section{Introduction}
\label{sec:introduction}

Forecasting outcomes in mixed-motive negotiations constitutes a substantial computational challenge, arising from the nonlinear interaction between overt linguistic communication and latent strategic constraints. In bargaining contexts, agents must simultaneously manage cooperative information exchange and competitive value appropriation. Their behavior is governed not only by explicit verbal statements but also by unobserved game-theoretic parameters, including budget limits, the Best Alternative to a Negotiated Agreement (BATNA), and Social Value Orientation (SVO)~\cite{chawla2021casino,fisher2011getting,murphy2011measuring}.

Prevailing computational approaches typically capture only one facet of this dual structure. Natural language processing models generally conceptualize negotiation as a sequence modeling or text classification problem, abstracting away the underlying game state and treating offers, threats, or bluffs merely as semantic variants~\cite{lewis2017deal,kwon2024llms}. Even state-of-the-art transformer-based agents exhibit limitations in agreement tracking, that is, the continuous verification of deal feasibility within a constrained solution space~\cite{mannekote2023agreement}. In contrast, classical game-theoretic agents impose strict mathematical constraints but largely ignore persuasive tactics, sentiment trajectories, and ambiguous communicative acts, which renders them brittle and poorly calibrated to realistic interactional settings~\cite{wasfy1998two}.

A central open problem is the principled integration of these heterogeneous modalities. Recent work has examined affective dynamics~\cite{lin2023toward} and pragmatic–semantic update mechanisms~\cite{petukhova2017computing}; however, standard fusion techniques (e.g., static feature concatenation) implicitly assume that modality salience remains constant across tasks and negotiation phases~\cite{qin2021co}. In practice, the relative informativeness of linguistic signals versus structural constraints is highly contingent on the negotiation topology: free-form dialogues depend predominantly on nuanced communicative cues, whereas structured bargaining tasks require strict adherence to game-theoretic feasibility constraints~\cite{he2021dialograph}. Existing architectures generally lack mechanisms to adaptively reweight these information sources as a function of negotiation structure and progression.

Furthermore, predictive models trained on observational data may encode and amplify distributive asymmetries present in the underlying datasets~\cite{kwon2024llms}. Conventional supervised learning objectives emphasize predictive accuracy or aggregate social welfare but rarely incorporate explicit criteria for the distribution of outcomes across agents. In this work, we adopt a reflective stance toward disparity modeling: instead of enforcing normative equality constraints, the proposed framework penalizes deviations from empirically observed utility gaps. This design choice aims to more faithfully reproduce the structural characteristics of individual negotiation instances, thereby enabling transparent and diagnostically useful analyses in decision-support and policy-evaluation contexts.

To address these limitations, this paper pursues the following objectives:
\begin{enumerate}
  \item To propose a unified dual-stream architecture that adaptively fuses semantic (textual) and strategic (graph-encoded economic state) modalities, enabling dynamic weighting of linguistic and game-theoretic signals.
  \item To introduce explicit grounding of latent constraints (e.g., BATNA, budgets, SVO-derived trust) into graph node features, overcoming limitations of text-only models in respecting rigid mathematical bounds.
  \item To incorporate a fairness-regularized composite loss that penalizes deviations from ground-truth utility disparities, thereby supporting reflective analysis of distributive outcomes.
  \item To empirically validate the framework's adaptability across contrasting negotiation environments (linguistic-heavy versus strategy-heavy), demonstrating effective joint prediction of deal occurrence and per-agent utility.
\end{enumerate}

The proposed Semantic-Temporal Graph Fusion Network (ST-GFN) consolidates these components into a unified computational architecture designed to support Group Decision and Negotiation Support Systems (GDNSS). By integrating adaptive multimodal data fusion with reflective disparity modeling, ST-GFN offers a theoretically grounded methodology for outcome prediction that anchors its inferences in both observable communicative behavior and underlying economic structures, thereby enabling more transparent, systematic, and equity-oriented analyses of negotiation dynamics.

\section{Related Work}

Foundational research in computational negotiation has predominantly focused on the processing of linguistic signals for the purpose of predicting negotiation outcomes. Chawla et al.~\cite{chawla2020exploring} introduced one of the earliest benchmarks on the CraigslistBargain dataset, demonstrating that BERT-based architectures can predict final agreed prices with an error margin of approximately 10\% using only partial dialogue transcripts, thereby outperforming LSTM-based baselines by roughly 10--20\%. Similarly, on the CaSiNo corpus, fine-tuned transformer models have been employed to classify persuasion strategies, achieving F1-scores in the range of 52\% to 68\% when combined with data augmentation techniques~\cite{chawla2021casino}.

More recent studies have leveraged Large Language Models (LLMs) to expand these capabilities. Kwon et al.~\cite{kwon2024llms} evaluated GPT-4 and Claude-3 on augmented CaSiNo tasks and reported 60--80\% accuracy on relatively shallow comprehension tasks. In contrast, performance declined markedly (to 40--60\%) on theory-of-mind inference tasks, highlighting the limitations of predominantly semantic processing for identifying latent intentions or deceptive behavior in the absence of explicit strategic priors. This limitation is further corroborated by Chan et al.’s NegotiationToM benchmark, which indicates that even state-of-the-art LLMs underperform humans by 20--30 percentage points in inferring interlocutors’ beliefs and intentions~\cite{chan2024negotiationtom}. Taken together, these findings suggest that although contemporary NLP models exhibit strong semantic processing capabilities, they remain vulnerable to textual noise and fail to fully capture the relational and strategic dynamics that are essential for high-stakes negotiation contexts.

To address the limitations of unimodal, text-only analysis, recent research has increasingly incorporated both strategic and affective signals via multimodal fusion architectures. Within the broader domain of dialogue modeling, methods such as Co-GAT leverage graph neural networks (GNNs) to perform joint dialogue act recognition and sentiment classification, yielding performance gains of approximately 5–10\% over unimodal baselines~\cite{qin2021co}. Similarly, GraphMFT employs multiple Graph Attention Networks (GATs) to conduct intra- and inter-modal fusion for emotion recognition, thereby enabling more effective exploitation of contextual dependencies~\cite{li2023graphmft}.

Graph-based approaches have proven particularly effective for modeling the relational structure that characterizes negotiation interactions. For instance, Bolleddu et al. integrate Graph Attention Networks with multi-agent reinforcement learning and report agreement rates of 94\% in simulated consensus tasks, substantially surpassing non-graph-based baselines~\cite{bolleddu2025dialogue}. However, many existing multimodal fusion techniques continue to rely on static concatenation or early-fusion paradigms, which are susceptible to propagating and amplifying noise from irrelevant or weakly informative modalities~\cite{gong2024multimodal}. The design and deployment of adaptive, gated fusion mechanisms, capable of dynamically reweighting the contributions of semantic and strategic signals across different negotiation phases, remain comparatively underexplored in the current literature.

Although contemporary NLP models exhibit strong capabilities for modeling communicative intent, effective negotiation further requires rigorous compliance with game-theoretic constraints, including maintaining a best alternative to a negotiated agreement (BATNA) and adhering to reservation prices.

Classical automated negotiating agents have predominantly employed heuristic search procedures or logic-based reasoning frameworks to optimize expected utility under explicit, hand-specified constraints~\cite{lin2014genius}. By contrast, contemporary deep learning–based approaches often struggle to reliably encode such rigid mathematical constraints when trained solely on textual corpora, which in turn yields strategically suboptimal concession dynamics~\cite{lewis2017deal}. 

Recent neuro-symbolic methods seek to embed these structural priors directly into neural architectures. For instance, Joshi et al. (DialoGraph) show that integrating a ``strategy graph'', which explicitly represents dependencies between sequences of pragmatic strategies and dialogue acts, improves both the coherence and the strategic consistency of generated responses~\cite{he2021dialograph}. Recent surveys of automated negotiation further emphasize the importance of hybrid symbolic–neural approaches to jointly capture linguistic and strategic reasoning~\cite{Beam1997AutomatedNA}. Extending this line of research, the proposed ST-GFN broadens the representational space to encode not only dialogue structure but also the underlying economic state (e.g., preference orderings, scalar BATNA values) within the graph features. This design grounds the neural model explicitly in the game state, thereby mitigating failure modes observed in text-only systems, which frequently hallucinate infeasible or economically inconsistent outcomes.

Despite advances in predictive performance, the systematic integration of fairness safeguards into negotiation-oriented AI remains nascent. Gratch~\cite{gratch2021promise} highlights the risks associated with automated negotiating agents, including deceptive conduct and moral distancing, and advocates incorporating ethical priors to counter these behaviors. Similarly, Iansiti and Lakhani argue that prevailing systems tend to reproduce and even exacerbate historical inequities present in their training data~\cite{iansiti2020competing}.

Empirical studies in applied contexts have documented biases that systematically advantage corporate or institutional stakeholders over individual parties in AI-assisted contract negotiation, with substantial consequences for distributive equity~\cite{snell2026artificial}. Although some recent work has started to incorporate fairness-aware reward formulations—for example, Bolleddu et al. report a 43\% reduction in inter-party utility disparity in simulation experiments~\cite{bolleddu2025dialogue}, most state-of-the-art models still prioritize strict utility maximization as their primary design criterion. In particular, LLM-based agents deployed in the NegotiationArena framework frequently exhibit aggression biases aimed at maximizing payoffs, explicitly trading off fairness for efficiency~\cite{bianchi2024well}. Recent advances in multi-agent fairness further underscore the necessity of explicitly disparity-aware objectives in negotiation environments~\cite{ju2024achieving}.

The extant literature reveals a marked methodological bifurcation: natural language processing techniques predominantly govern semantic representation and analysis, whereas emerging graph-based methodologies primarily enhance relational and structural inference. However, there is currently no unified framework that adaptively integrates these modalities while explicitly imposing reflective constraints on distributive outcomes. Existing multimodal architectures frequently attain superior predictive performance but concurrently risk perpetuating and exacerbating historical biases embedded in benchmark datasets such as CaSiNo. The proposed ST-GFN framework seeks to mitigate this limitation by introducing a gated fusion mechanism that dynamically reweights semantic and strategic feature representations, complemented by a fairness-regularized loss function designed to promote reflectively constrained utility prediction across interacting parties.

\section{Methodology}

To systematically address the challenges associated with modeling complex, dynamic, and multi-party mixed-motive negotiations, we conceptualize the negotiation environment as a dual-stream computational framework that encompasses the entire process of group decision-making and negotiation. Formally, a negotiation instance is represented as the tuple $\mathcal{N} = (\mathcal{X}_{\text{txt}}, \mathcal{G}_{\text{strat}}, \mathbf{y})$, where $\mathcal{X}_{\text{txt}}$ denotes the sequence of linguistic interactions, $\mathcal{G}_{\text{strat}}$ encodes the underlying strategic structure and economic constraints, and $\mathbf{y}$ represents the negotiation outcomes, including agreement status and agent-specific utilities. The primary objective is to learn a parameterized mapping
\begin{equation}
    f_\theta: (\mathcal{X}_{\text{txt}}, \mathcal{G}_{\text{strat}}) \rightarrow \mathbf{y}
\end{equation}
that not only approximates negotiation outcomes with high predictive accuracy but also facilitates systematic analysis of the distributive and procedural dynamics that characterize group decision-making processes.

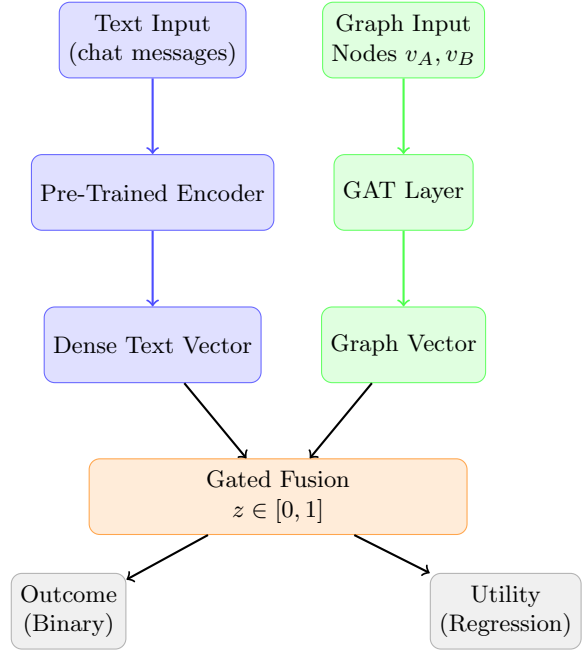
\begin{figure}[h!]
\centering
\begin{tikzpicture}[
    node distance=1 cm and 1 cm,
    every node/.style={
        draw,
        rounded corners,
        align=center,
        minimum height=1cm,
        font=\small
    },
    textnode/.style={fill=blue!12, draw=blue!60},
    graphnode/.style={fill=green!12, draw=green!60},
    fusionnode/.style={fill=orange!15, draw=orange!70, minimum width=5cm},
    outnode/.style={fill=gray!12, draw=gray!60},
    arrow/.style={->, thick}
]
\node[textnode] (text_input) {Text Input\\(chat messages)};
\node[textnode] (roberta) [below=of text_input] {Pre-Trained Encoder};
\node[textnode] (text_vec) [below=of roberta] {Dense Text Vector};
\node[graphnode] (graph_input) [right=1 cm of text_input] {Graph Input\\Nodes $v_A, v_B$};
\node[graphnode] (gat) [below=of graph_input] {GAT Layer};
\node[graphnode] (graph_vec) [below=of gat] {Graph Vector};
\node[fusionnode] (fusion) [below=1.5 cm of $(text_vec)!0.5!(graph_vec)$] {Gated Fusion\\$z \in [0,1]$};
\node[outnode] (outcome) [below left=.5 cm and -.5 cm of fusion] {Outcome\\(Binary)};
\node[outnode] (utility) [below right=.5 cm and -0.5 cm of fusion] {Utility\\(Regression)};
\draw[arrow, blue!70] (text_input) -- (roberta);
\draw[arrow, blue!70] (roberta) -- (text_vec);
\draw[arrow] (text_vec) -- (fusion);
\draw[arrow, green!70] (graph_input) -- (gat);
\draw[arrow, green!70] (gat) -- (graph_vec);
\draw[arrow] (graph_vec) -- (fusion);
\draw[arrow] (fusion) -- (outcome);
\draw[arrow] (fusion) -- (utility);
\end{tikzpicture}
\caption{The ST-GFN Dual-Stream Architecture. The model processes the negotiation tuple $\mathcal{N}$ via two parallel modalities. The Linguistic Stream (left) maps dialogue turns $T$ to semantic vectors $\mathbf{H}_{txt}$, while the Strategic Stream (right) models agent constraints as a directed graph $\mathcal{G}$. These streams are dynamically fused via a learned gate $z \in [0,1]$ to support prediction of both agreement occurrence and distributive outcomes in group decision and negotiation processes.}
\label{fig:architecture}
\end{figure}

The input space consists of two complementary modalities that reflect the dual nature of real-world negotiations: manifest linguistic interaction and latent strategic structure as shown in Figure~\ref{fig:architecture}. The linguistic modality $\mathcal{X}_{\text{txt}}$ represents the dialogue as an ordered sequence of turns $T = \{t_1, t_2, \dots, t_K\}$, where each turn $t_k$ is mapped into a high-dimensional embedding space by a pre-trained transformer encoder. This procedure yields a sequence of semantic representations $\mathbf{H}_{\text{txt}} \in \mathbb{R}^{K \times d_{\text{txt}}}$ that encode persuasion tactics, temporal sentiment dynamics, offers, counteroffers, acceptance cues, and other communicative acts that are central to the information-exchange and problem-formulation phases of negotiation.

In parallel, we construct a directed graph $\mathcal{G} = (\mathcal{V}, \mathcal{E})$ to explicitly model the dyadic power relations and interactional dependencies that shape the feasible outcome space. The node set $\mathcal{V} = \{v_A, v_B\}$ corresponds to the two negotiating parties.

Each node embedding $v_i \in \mathbb{R}^{d_{\text{node}}}$ parameterizes a collection of scalar features that encode the agent’s economic status, strategic constraints, and behavioral dispositions. These features typically comprise quantities such as BATNA, budget or resource limitations, role descriptors, SVO, priority weights over negotiation issues, and other domain-specific metadata. The precise specification and estimation of these attributes are conditioned on the information available within the negotiation corpus, thereby ensuring that the resulting graph representation remains aligned with the structural properties of the underlying environment. 

The edge set $\mathcal{E}$ represents inter-agent relationships and is parameterized by a Trust Score $\tau_{ij}$ for each ordered pair of agents. This score is typically derived from the agents’ SVO or other relational indicators, thereby incorporating psychological predispositions toward cooperation or competition directly into the graph’s weighted topology. Such relational encoding is particularly valuable for negotiation, as it allows the framework to model how trust and power asymmetries evolve during the negotiation process.

The central component of the proposed framework is ST-GFN, illustrated in Figure~\ref{fig:architecture}. The design is motivated by the observation that negotiation outcomes are determined by the interaction between explicit linguistic signals and latent strategic constraints, which are often decoupled as a consequence of deliberate misrepresentation or bluffing. The architecture jointly models these elements to provide a more comprehensive foundation for GDNSS.

The strategic interaction graph is initially encoded using a GAT, which refines agent representations as a function of their neighbors’ states and the corresponding trust levels:
\begin{equation}
    \mathbf{h}_{\text{graph}} = \text{GAT}(\mathcal{V}, \mathcal{E}, \tau).
\end{equation}

To exploit the complementary, non-redundant information contained in both modalities, we introduce a gated multimodal fusion mechanism. The textual and graph-based embeddings are concatenated to form a joint representation $\mathbf{c}_k = [\mathbf{h}_{\text{txt}}^{(k)} \parallel \mathbf{h}_{\text{graph}}]$. A learnable fusion gate $z_k \in (0, 1)$ is then computed via a sigmoid activation:
\begin{equation}
    z_k = \sigma(\mathbf{W}_z \mathbf{c}_k + b_z).
\end{equation}
The final fused representation is defined as:
\begin{equation}
    \mathbf{h}_{\text{fused}}^{(k)} = z_k \odot \text{ReLU}(\mathbf{W}_p \mathbf{c}_k + b_p).
\end{equation}

\begin{figure}[h!]
    \centering
    \includegraphics[width=0.6\linewidth]{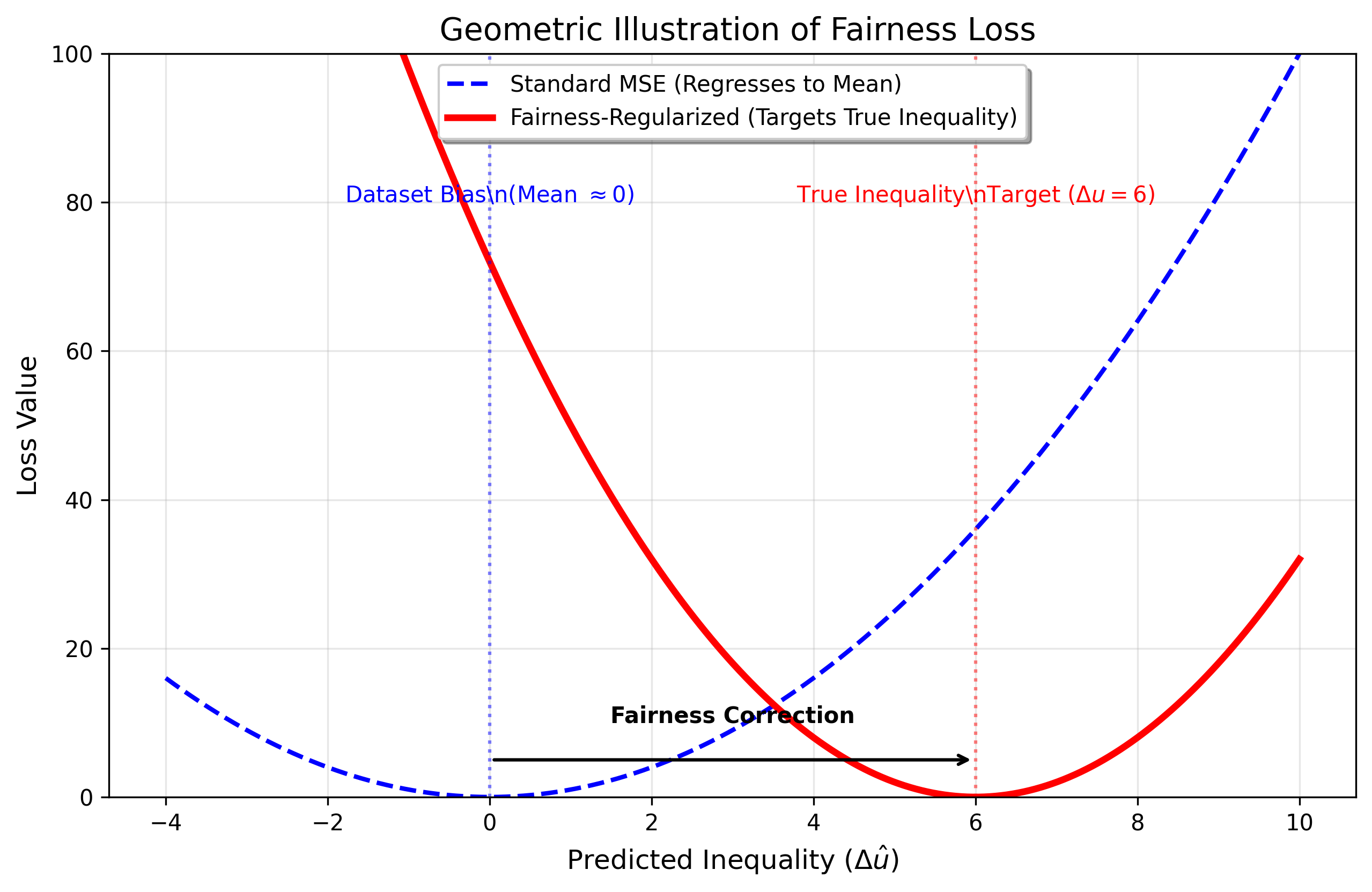}
    \caption{\textbf{Geometric Interpretation of Fairness Loss.} The dashed blue curve represents a standard MSE objective, which minimizes error towards the dataset mean (zero inequality). The solid red curve illustrates our Fairness-Regularized objective. By anchoring the loss minimum to the true inequality target ($\Delta u=6$) rather than the global average, the function forces the model to recognize and predict structural imbalances inherent in the negotiation.}
    \label{fig:fairness_loss}
\end{figure}

This gating mechanism enables the model to adaptively reweight semantic information relative to strategic constraints at each timestep, thereby accommodating the heterogeneous informational requirements that arise across distinct phases and types of negotiation.

The resulting sequence of fused vectors $\{\mathbf{h}_{\text{fused}}^{(k)}\}_k$ is subsequently processed by a Long Short-Term Memory (LSTM) network in order to explicitly model the temporal dynamics of the negotiation process. This form of sequential modeling is critical for capturing turn-by-turn variations in bargaining power, concession trajectories, sentiment dynamics, and the progressive convergence (or divergence) toward agreement, core phenomena in the empirical and computational study of group decision-making and negotiation processes.

A central contribution of this work is the explicit treatment of fairness and distributional properties of negotiation outcomes. Conventional supervised learning objectives prioritize predictive accuracy or aggregate utility, typically without regard to how benefits and burdens are allocated across negotiating parties. In negotiation settings, such neglect can obscure underlying structural inequalities. 

To facilitate normative and reflective evaluation, we optimize a composite objective function
\begin{equation}
\mathcal{L} = \mathcal{L}_{\text{outcome}} + \mathcal{L}_{\text{utility}} + \lambda \mathcal{L}_{\text{fairness}},
\end{equation}
where $\mathcal{L}_{\text{outcome}}$ and $\mathcal{L}_{\text{utility}}$ correspond to the standard losses for deal outcome classification and utility regression, respectively. The fairness regularization component is given by
\begin{equation}
   \mathcal{L}_{\text{fairness}} = \left( \Delta(\hat{\mathbf{u}}) - \Delta(\mathbf{u}) \right)^2,
\end{equation}
with $\Delta(\mathbf{u}) = |u_A - u_B|$ denoting the absolute utility difference between the two parties. This term incentivizes the model to faithfully reproduce the realized disparity in utilities for each individual negotiation instance, rather than collapsing toward an average notion of equity. As illustrated in Figure~\ref{fig:fairness_loss}, this design anchors predictions in the empirical power relations that characterize the interaction, thereby enabling transparent assessment of distributive justice without enforcing an exogenous egalitarian bias.

The complete training procedure is formalized in Algorithm~\ref{alg:training}. By integrating the inequality penalty directly into backpropagation, the framework jointly optimizes predictive performance and the faithful representation of distributive outcomes, offering a design-oriented contribution to the development of ethically reflective group decision and negotiation.

\begin{algorithm}[h!]
\caption{Training Procedure for ST-GFN with Fairness Constraints}
\label{alg:training}
\begin{algorithmic}[1]
\REQUIRE Dataset $\mathcal{D} = \{(\mathcal{X}_{\text{txt}}^{(i)}, \mathcal{G}_{\text{strat}}^{(i)}, \mathbf{y}^{(i)})\}_{i=1}^N$, Fairness weight $\lambda$, Learning rate $\eta$
\ENSURE Trained parameters $\theta$
\STATE Initialize model parameters $\theta$ (Encoder, GAT, LSTM, Gates)
\FOR{epoch $= 1$ to $E$}
    \FOR{batch $B \subset \mathcal{D}$}
        \STATE \textbf{1. Dual-Stream Encoding}
        \STATE $\mathbf{H}_{\text{txt}} \leftarrow \text{Pre-Trained Encoder}(B.\mathcal{X}_{\text{txt}})$ \COMMENT{Semantic embeddings}
        \STATE $\mathbf{h}_{\text{graph}} \leftarrow \text{GAT}(B.\mathcal{G}_{\text{strat}}, \tau)$ \COMMENT{Strategic embeddings with Trust $\tau$}
        \STATE \textbf{2. Gated Fusion}
        \STATE $\mathbf{c} \leftarrow [\mathbf{H}_{\text{txt}} \parallel \mathbf{h}_{\text{graph}}]$ \COMMENT{Concatenation}
        \STATE $z \leftarrow \sigma(\mathbf{W}_z \mathbf{c} + b_z)$ \COMMENT{Compute Gate}
        \STATE $\mathbf{h}_{\text{fused}} \leftarrow z \odot \text{ReLU}(\mathbf{W}_p \mathbf{c} + b_p)$ \COMMENT{Apply Fusion}
        \STATE \textbf{3. Prediction}
        \STATE $\hat{y}_{\text{outcome}}, \hat{\mathbf{u}}_{\text{utility}} \leftarrow \text{LSTM}(\mathbf{h}_{\text{fused}})$
        \STATE \textbf{4. Fairness-Aware Loss Calculation}
        \STATE $\mathcal{L}_{\text{outcome}} \leftarrow \text{BCE}(\hat{y}_{\text{outcome}}, y_{\text{outcome}})$
        \STATE $\mathcal{L}_{\text{utility}} \leftarrow \text{MSE}(\hat{\mathbf{u}}_{\text{utility}}, \mathbf{u}_{\text{utility}})$
        \STATE $\Delta_{\text{pred}} \leftarrow |\hat{u}_A - \hat{u}_B|$; \quad $\Delta_{\text{true}} \leftarrow |u_A - u_B|$
        \STATE $\mathcal{L}_{\text{fairness}} \leftarrow (\Delta_{\text{pred}} - \Delta_{\text{true}})^2$ \COMMENT{Inequality Penalty}
        \STATE $\mathcal{L}_{\text{total}} \leftarrow \mathcal{L}_{\text{outcome}} + \mathcal{L}_{\text{utility}} + \lambda \mathcal{L}_{\text{fairness}}$
        \STATE \textbf{5. Optimization}
        \STATE $\theta \leftarrow \theta - \eta \nabla_\theta \mathcal{L}_{\text{total}}$ \COMMENT{Backpropagation}
    \ENDFOR
\ENDFOR
\RETURN $\theta$
\end{algorithmic}
\end{algorithm}

Algorithm \ref{alg:training} formalizes the iterative optimization procedure underlying the ST-GFN framework. The process begins with the dual-stream encoding stage (Lines 4–6), in which unstructured dialogue turns are embedded via a Transformer-based encoder, while strategic states are represented using a Graph Attention Network (GAT). A central methodological contribution is the gated fusion mechanism (Lines 7–10), wherein the fusion gate $z$ functions as a differentiable selector that adaptively routes gradients toward either the semantic or strategic pathway, contingent on their relative predictive utility at each timestep.

Subsequently, the optimization objectives are evaluated within the fairness-aware loss module (Lines 13–17). In contrast to conventional architectures that decouple utility estimation from disparity assessment, the proposed algorithm embeds the inequality penalty $ \mathcal{L}_{\text{fairness}} $ directly into the backpropagation process. This integration enforces a regularizing constraint on the parameter space $ \theta $, thereby steering gradient descent away from local minima that reduce prediction error through trivial mean aggregation (i.e., convergence to uniformly equal predictions for intrinsically heterogeneous inputs).

\section{Experimental Analysis}

To evaluate the contributions of the ST-GFN framework to GDNSS, we conducted experiments that assess its predictive performance, cross-topology adaptability, temporal fusion dynamics, and capacity to reflect distributive outcomes. The analysis demonstrates how the dual-stream architecture and reflective regularization support key GDNSS functions: monitoring evolving bargaining power, grounding predictions in strategic constraints, and enabling transparent assessment of equity in negotiation processes.

\subsection{Dataset and Pre-processing}
We selected two benchmark datasets that instantiate distinct negotiation structures to evaluate the robustness and generalization capabilities of the proposed framework: the strategy-centric CaSiNo corpus and the linguistically oriented DealOrNoDeal corpus. This selection enables a controlled examination of how ST-GFN adapts to varying emphases on linguistic persuasion versus strategic constraints, in line with prior GDNSS research on heterogeneous multiplayer interaction settings.

The CaSiNo dataset comprises 1,030 bilateral negotiation dialogues \cite{chawla2021casino}. In each dialogue, two agents (``Camp Manager'' and ``Unit Ranger'') negotiate the allocation of three resource types: food, water, and firewood. The corpus provides rich structured metadata, including role-specific constraints, private BATNA values, and ordinal priority hierarchies (High/Medium/Low). These annotations support detailed analyses of both game-theoretic properties and relational dynamics.

The DealOrNoDeal dataset contains 5,808 dialogues focused on allocating discrete items (books, hats, and balls) \cite{lewis2017deal}. In contrast to CaSiNo, it does not provide explicit role-based metadata; instead, each agent is assigned randomized scalar valuations (0–10) for each item type. This design emphasizes evaluation in comparatively unconstrained, communication-driven environments.

Taken together, these contrasting datasets facilitate a systematic assessment of ST-GFN’s capacity to adaptively support diverse negotiation topologies within GDNSS scenarios, ranging from communication-centric persuasion to more tightly constrained bargaining settings.

For both datasets, raw dialogue turns were tokenized using the RoBERTa-base tokenizer \cite{zhuang-etal-2021-robustly}, with a maximum sequence length of $L = 128$. The strategic graph $ \mathcal{G}_{\text{strat}} $ was instantiated using node features tailored to the available metadata while remaining consistent with the methodology’s general design. Specifically, we incorporated economic parameters (e.g., BATNA, budgetary constraints), relational indicators (e.g., Trust Score derived from SVO annotations), and issue priorities encoded as numerical weights where applicable.

Preliminary dataset analysis revealed a pronounced class imbalance, with successful agreements ($y = 1$) substantially more frequent than failures. To reduce bias toward majority-class predictions, we applied minority-class oversampling during training by sampling ``No-Deal'' instances with replacement until achieving an approximately balanced (1:1) class ratio. The validation and test sets preserved the original label distributions to ensure realistic performance evaluation.

\subsection{Implementation Details}
The framework was implemented in PyTorch \cite{10.5555/3454287.3455008}, employing PyTorch Geometric for graph-based operations \cite{fey2019fast}. All experiments were conducted on an Apple M-series processor using Metal Performance Shaders for hardware acceleration. Reported results were averaged over five random seeds (42–46) to enhance statistical robustness.

Model training utilized the AdamW optimizer with a weight decay of $1 \times 10^{-4}$ for a maximum of 100 epochs, together with a \texttt{ReduceLROnPlateau} learning-rate scheduler monitoring the validation loss (patience of 5 epochs; decay factor of 0.1). Textual representations (768-dimensional) were obtained from a frozen RoBERTa-base model and remained fixed during training. The full set of hyperparameters is provided in Table \ref{tab:hyperparams}. A fairness coefficient of $\lambda = 0.7$ was employed to balance predictive performance and reflective regularization.

The evaluation protocol comprises Accuracy, F1-score, and AUC-ROC for outcome classification; MAE and MSE for utility regression; and Inequality Discrepancy (ID) to quantify the fidelity with which the model captures distributive outcomes:
$$
\text{ID} = \frac{1}{N} \sum_{i=1}^{N} \left| \,\bigl|\hat{u}^{(i)}_1 - \hat{u}^{(i)}_2\bigr| - \bigl|u^{(i)}_1 - u^{(i)}_2\bigr| \right|.
$$
Lower ID values correspond to more accurate modeling of inter-agent utility differentials, thereby supporting normative analysis within GDNSS. The metrics were selected to assess not only standard classification and regression accuracy but also the capacity of the framework to faithfully represent disparities between agents—an essential prerequisite for rigorous normative evaluation and equity-aware decision support in GDNSS.

\begin{table}[h!]
\centering
\caption{Hyperparameter settings for the ST-GFN experiments.}
\label{tab:hyperparams}
\small
\setlength{\tabcolsep}{4pt}

\begin{tabularx}{\columnwidth}{l l c X}
\toprule
\textbf{Category} & \textbf{Parameter} & \textbf{Value} & \textbf{Notes} \\
\midrule

\textbf{Optimization} & Batch Size & 16 & Mini-batch training \\
 & Learning Rate ($\eta$) & $1 \times 10^{-4}$ & AdamW base LR \\
 & Optimizer & AdamW & Weight decay enabled \\
 & Max Epochs & 100 & Full training budget \\
 & Scheduler Patience & 5 epochs & Early LR decay \\
 & Fairness Weight ($\lambda$) & 0.7 & Fairness–performance tradeoff \\
 & Random Seeds ($N$) & 5 & Averaged runs \\

\midrule

\textbf{Architecture} & Text Embedding Dim & 768 & Frozen RoBERTa encoder \\
 & Node Feature Dim & 6 & Structural attributes \\
 & Hidden Dimension ($d$) & 128 & GNN hidden size \\
 & Dropout Rate & 0.3 & Regularization \\
 & GAT Attention Heads & 2 & Multi-head attention \\
 & Sequence Length (Turns) & 10 & Dialogue horizon \\

\bottomrule
\end{tabularx}
\end{table}

\subsection{Performance Evaluation}
\label{perform}

We compared the ST-GFN framework against a logistic regression baseline and an unconstrained ablation ($\lambda=0$) to assess its utility for GDNSS. As summarized in Table \ref{tab:main_results}, the framework achieves strong outcome prediction while substantially enhancing reflection of distributive outcomes.

\begin{table*}[h]
\centering
\caption{Main Experimental Results ($N=5$ Seeds). The baseline achieves comparable accuracy but exhibits large errors in utility estimation and disparity reflection. ST-GFN with fairness regularization yields a 43.8\% reduction in Inequality Discrepancy with minimal impact on accuracy.}
\label{tab:main_results}
\resizebox{\textwidth}{!}{%
\begin{tabular}{l|cc|cc|c}
\hline
\textbf{Model Configuration} & \textbf{Accuracy} $\uparrow$ & \textbf{F1-Score} $\uparrow$ & \textbf{Utility MAE} $\downarrow$ & \textbf{Inequality Discrepancy} $\downarrow$ & \textbf{Reduction in ID} \\ \hline
Baseline (Logistic Reg.) & \textbf{0.9612} $\pm$ 0.009 & \textbf{0.9801} $\pm$ 0.005 & 11.43 $\pm$ 0.36 & 13.44 $\pm$ 1.14 & -- \\
ST-GFN (No Fairness) & 0.9563 $\pm$ 0.005 & 0.9776 $\pm$ 0.003 & \textbf{2.56} $\pm$ 0.11 & 3.79 $\pm$ 0.10 & -- \\
\textbf{ST-GFN (With Fairness)} & 0.9553 $\pm$ 0.007 & 0.9771 $\pm$ 0.004 & 2.90 $\pm$ 0.12 & \textbf{2.13} $\pm$ 0.13 & \textbf{43.8\%} \\ \hline
\end{tabular}%
}
\end{table*}

\begin{figure*}[t!]
\centering
\begin{subfigure}[b]{0.48\textwidth}
\centering
\includegraphics[width=\linewidth]{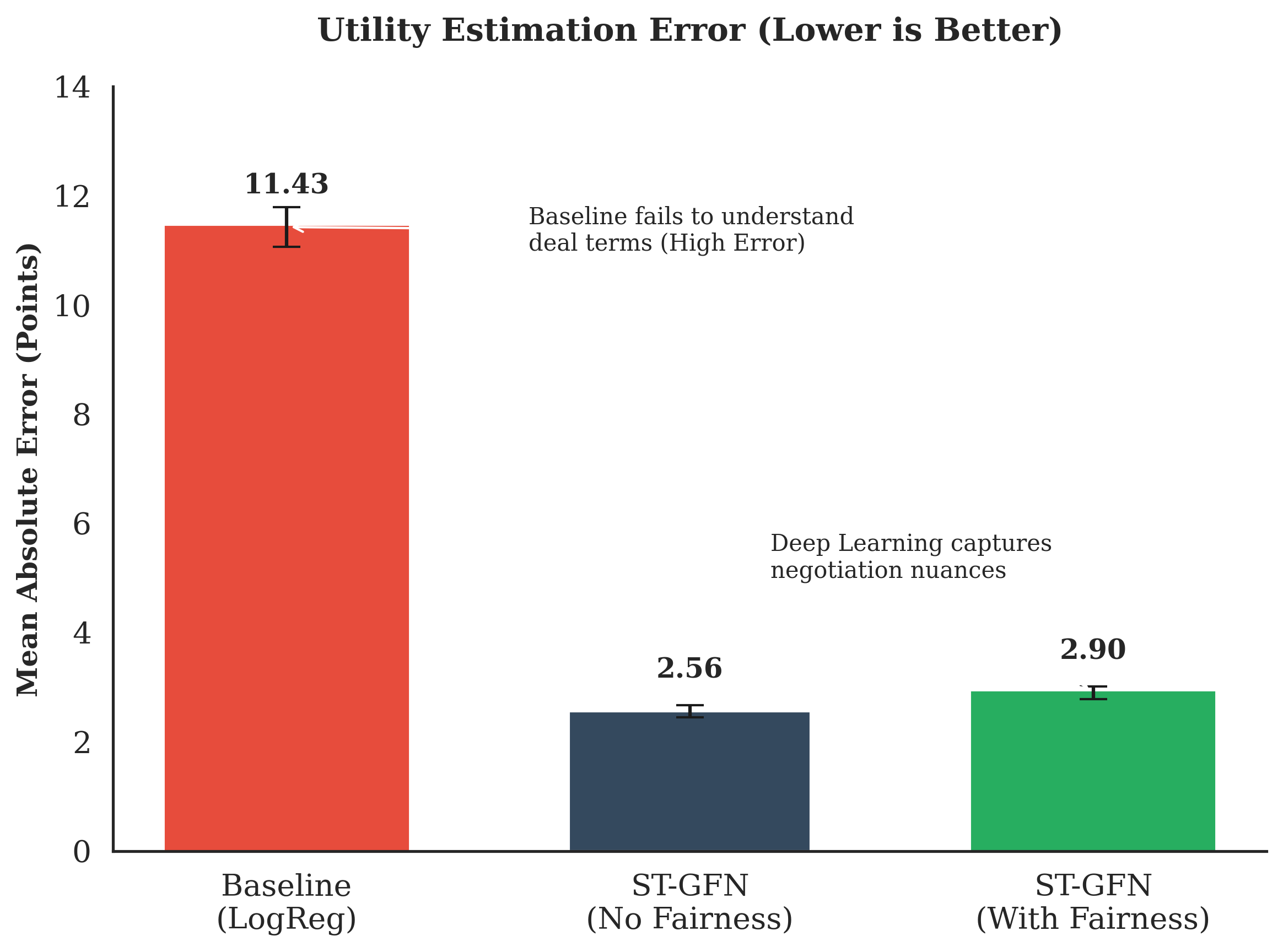}
\caption{Utility Estimation Error. The baseline (red) exhibits large errors, while ST-GFN (green) enables precise utility modeling essential for GDNSS feasibility assessment.}
\label{fig:utility_error}
\end{subfigure}
\hfill
\begin{subfigure}[b]{0.48\textwidth}
\centering
\includegraphics[width=\linewidth]{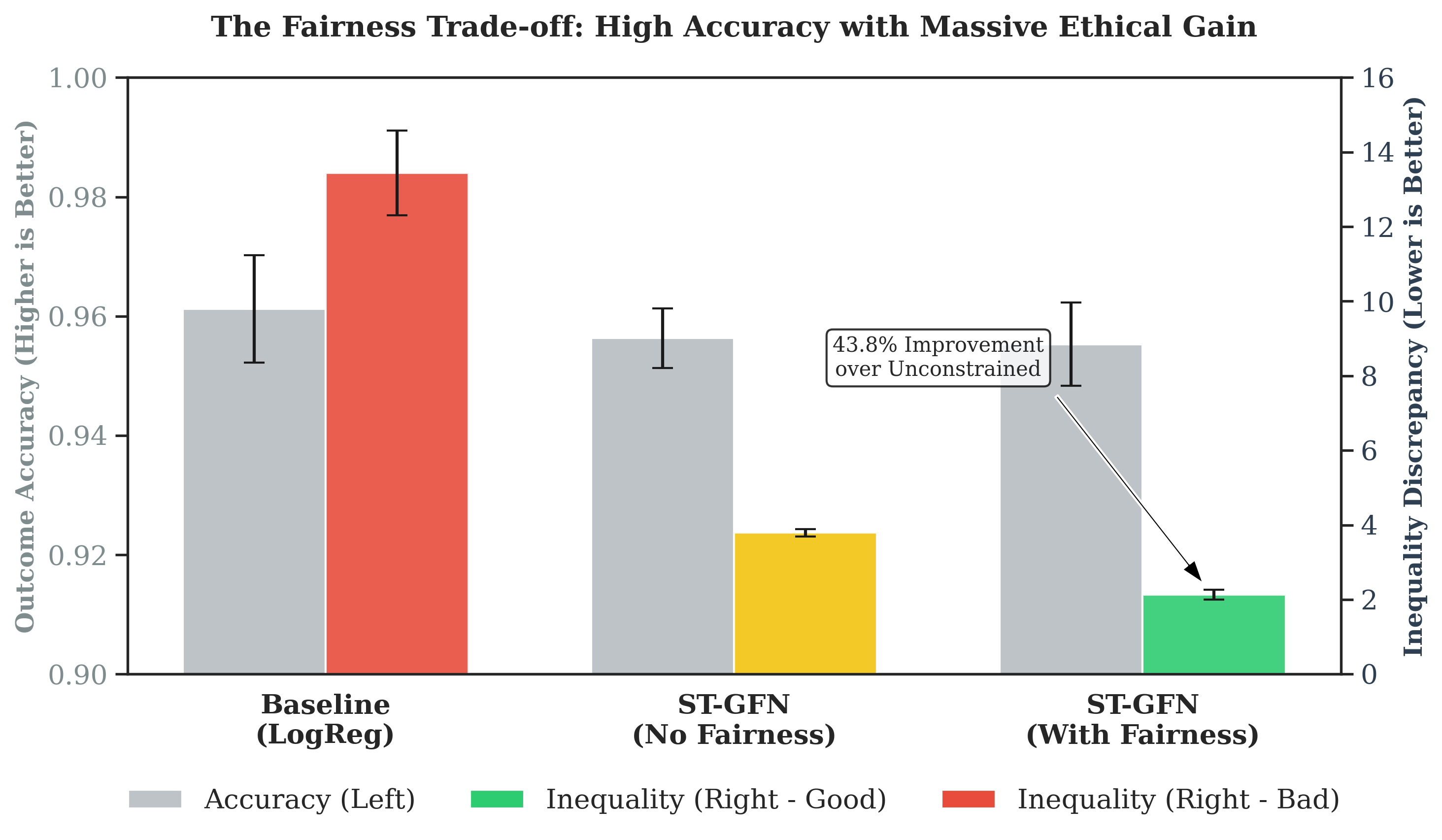}
\caption{Accuracy vs. Discrepancy Trade-off. Outcome accuracy (left axis) remains stable across models, while Inequality Discrepancy (right axis) decreases substantially with fairness regularization.}
\label{fig:fairness_tradeoff}
\end{subfigure}
\caption{Performance and Disparity Analysis. (a) Comparison of Mean Absolute Error (MAE) in utility regression. (b) Dual-axis view of outcome accuracy and Inequality Discrepancy.}
\label{fig:combined_performance}
\end{figure*}

For binary outcome prediction, all models exceed 95\% accuracy, with no statistically significant differences (two-tailed Wilcoxon signed-rank test, $p=0.250$). This indicates that linguistic markers alone are often sufficient for detecting agreement in these datasets.

The substantial reduction in utility MAE (from 11.43 to 2.90) demonstrates ST-GFN’s ability to model fine-grained concession patterns and value exchanges, a prerequisite for real-time feasibility checking and concession guidance in GDNSS. Figure \ref{fig:utility_error} illustrates this performance gap, with ST-GFN supporting detailed utility analysis relevant to negotiation support.

The ablation study shows that reflective regularization imposes only a minor predictive cost (MAE increase from 2.56 to 2.90) while delivering a 43.8\% improvement in disparity reflection (Inequality Discrepancy reduced from 3.79 to 2.13). Figure \ref{fig:fairness_tradeoff} confirms stable outcome accuracy alongside substantial gains in equitable representation. These results support the design of GDNSS tools that can alert negotiators to structural imbalances without compromising core predictive reliability.

Overall, these findings indicate that ST-GFN can serve as a computational backbone for GDNSS applications requiring simultaneous monitoring of communicative flow and distributive equity.

\subsection{Temporal Dynamics of Multimodal Fusion}

To examine the adaptability of the gated fusion mechanism and its implications for dynamic negotiation support in GDNSS, we analyzed the temporal behavior of the gate parameter $z_k$ across the test set.

\begin{figure*}[h!]
\centering
\includegraphics[width=1.0\textwidth]{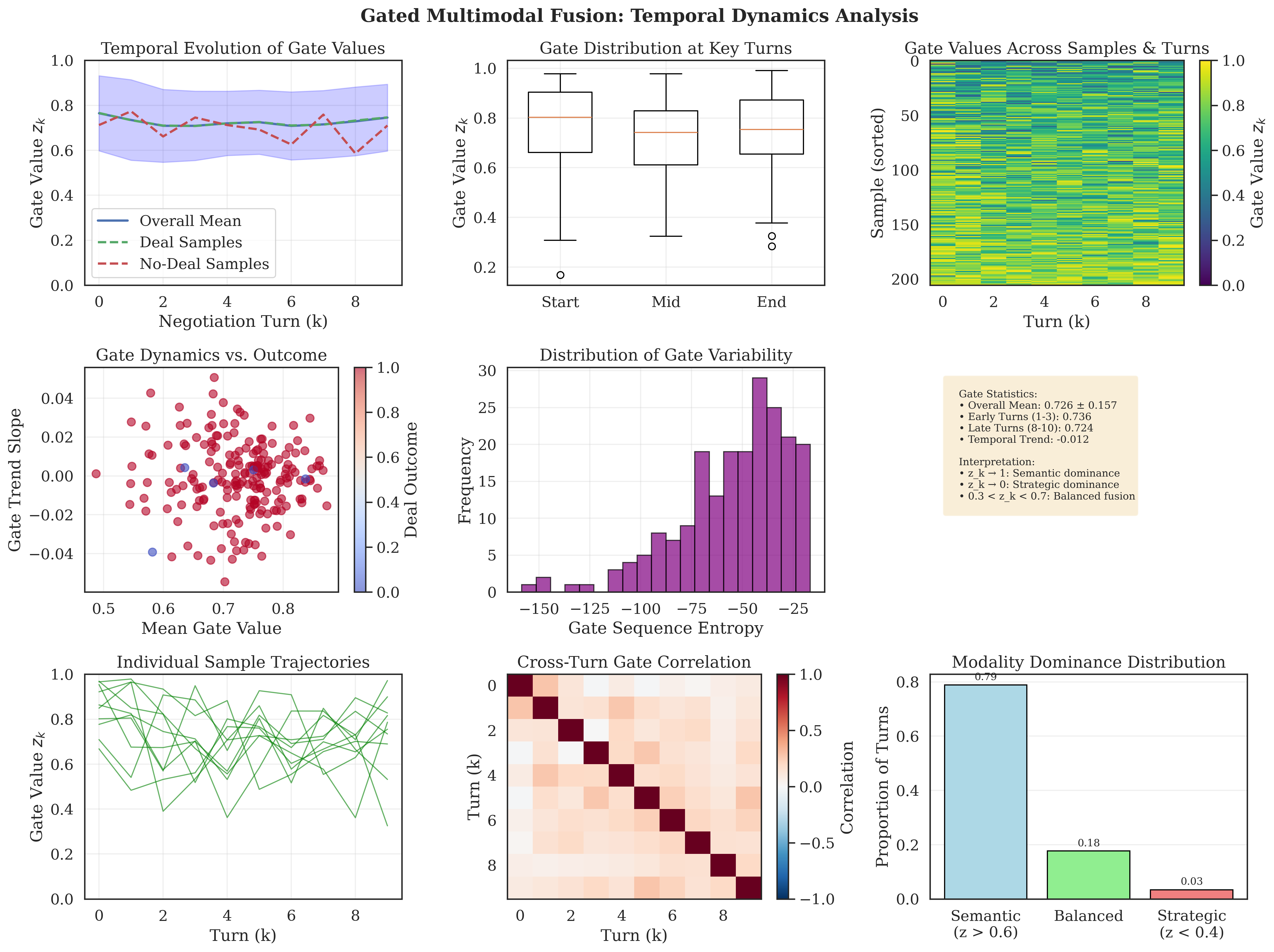}
\caption{Gated Fusion Dynamics Dashboard. The temporal evolution plot (top left) demonstrates overall stability. The heatmap of gate values across samples and turns (top right) reveals session-specific consistency rather than systematic temporal shifts. The modality dominance distribution (bottom right) highlights predominant linguistic contribution with strategic graph support for constraint enforcement.}
\label{fig:gate_dashboard}
\end{figure*}

Figure \ref{fig:gate_dashboard} illustrates temporal stability in the CaSiNo dataset ($\mu_z = 0.726 \pm 0.157$; linear slope $\approx -0.001$), indicating consistent hybrid weighting (approximately 73\% linguistic, 27\% strategic). This stable integration of communicative and structural signals supports GDNSS by grounding predictive inferences in both negotiation dialogue and latent constraints, enabling reliable monitoring of bargaining power across turns.

Individual gate trajectories exhibit considerable heterogeneity, with fusion strategies adapting to the specific characteristics of each negotiation session. Successful agreements display relatively stable gate profiles, whereas failed negotiations show greater volatility, consistent with heightened uncertainty and dynamic resolution of communicative ambiguity during the process.

The modality dominance distribution reveals that 79\% of turns are linguistically dominant ($z > 0.6$), while the strategic graph serves primarily as a constraint-checking mechanism (strict dominance $z < 0.4$ in only 3\% of turns). This distribution underscores the graph's complementary role in ensuring economic feasibility, thereby enhancing the framework's suitability for GDNSS applications that require adaptive, context-sensitive support.

\subsection{State-of-the-Art Comparison}

\begin{table*}[htbp]
\centering
\caption{Comparison with State-of-the-Art Frameworks. ST-GFN advances joint prediction of negotiation outcomes and utilities while incorporating reflective modeling of inter-agent utility disparities.}
\label{tab:sota_comparison}
\resizebox{\textwidth}{!}{%
\begin{tabular}{l p{3.5cm} p{4cm} p{5cm} p{3.5cm}}
\toprule
\textbf{Model/Framework} & \textbf{Dataset} & \textbf{Primary Task} & \textbf{Key Metrics} & \textbf{Disparity Mechanism} \\
\midrule
\textbf{ST-GFN (Proposed)} & CaSiNo & Joint Outcome \& Utility Prediction & Acc: 0.955; F1: 0.977; MAE: 2.90; ID: 2.13 & Reflective Loss: Penalizes deviation from true utility gaps. \\
\midrule
BERT-Negotiation \cite{chawla2020exploring} & CraigslistBargain & Price Prediction \& Strategy Labeling & Price prediction within 10\% margin for >70\% cases; Strategy F1 $\approx$ 68\% & None. \\
\midrule
NegotiationToM \cite{chan2024negotiationtom} & Augmented CaSiNo & Mental State Inference (Beliefs, Desires, Intentions) & Accuracy: 40--60\% on mental states & None. \\
\midrule
NegotiationArena \cite{bianchi2024well} & Synthetic Scenarios (e.g., Ultimatum games) & End-to-End Negotiation Simulation & Agreement Rate: 50--80\%; Payoff gains up to $\sim$20\% with tactics & Incidental (observes inherent biases in LLMs). \\
\midrule
LLM-Negotiator \cite{kwon2024llms} & Augmented CaSiNo & Negotiation Comprehension \& Simulation & Success Rate: 60--80\% (simple tasks); 40--60\% (complex tasks) & Partial: Notes fairness/aggression biases, but no explicit control mechanism. \\
\midrule
Zooming Out \cite{distasi2024zooming} & Lab-based Dialogues (Multi-issue) & Outcome \& Relational Prediction & $R^2 \approx 0.20$--0.40 for objective/relational outcomes & None. \\
\bottomrule
\end{tabular}%
}
\end{table*}

Table \ref{tab:sota_comparison} presents a comparative analysis of the proposed ST-GFN framework and representative state-of-the-art methodologies in computational negotiation modeling. The comparison focuses on the primary modeling objectives, core evaluation metrics, and the extent to which each approach explicitly represents and regulates inter-agent utility disparities.

The ST-GFN framework is characterized by its joint prediction of both negotiation outcomes and agent-specific utilities, augmented by a reflective regularization term that penalizes deviations from empirically observed utility gaps. This design enables faithful modeling of distributive dynamics without imposing artificial egalitarian constraints, thereby supporting more rigorous normative analysis and fairness-aware facilitation within GDNSS.

By contrast, earlier approaches such as BERT-Negotiation \cite{chawla2020exploring} and the original CaSiNo framework primarily target price prediction or strategic action labeling and do not incorporate mechanisms for modeling or controlling utility disparities. Subsequent extensions, including NegotiationToM \cite{chan2024negotiationtom} and LLM-based systems \cite{kwon2024llms, bianchi2024well}, emphasize mental-state inference, scenario simulation, or textual comprehension, but generally lack explicit strategic inductive biases or structured disparity regulation. Similarly, relational and affective modeling approaches \cite{distasi2024zooming} provide statistical characterizations of conversational and socio-emotional dynamics but do not embed formal game-theoretic foundations or fairness-oriented regularization objectives.

Overall, the comparison indicates that existing models tend to excel on isolated subtasks, yet none concurrently deliver joint outcome–utility prediction coupled with reflective disparity modeling. ST-GFN addresses this methodological gap through adaptive multimodal fusion and a fairness-aware training objective, thereby advancing the development of GDNSS tools that simultaneously enhance predictive accuracy and support equity-sensitive decision processes in negotiation settings.

\subsection{Cross-Dataset Analysis: Modality Adaptability}
To assess ST-GFN’s adaptability across heterogeneous negotiation topologies and its implications for flexible GDNSS deployment, we conduct a comparative evaluation on two benchmark datasets: the linguistically oriented DealOrNoDeal and the strategy-centric CaSiNo. As summarized in Table \ref{tab:cross_dataset}, the framework dynamically modulates modality-specific contributions while preserving strong predictive performance and improving the fidelity of disparity representation.

The mean gate value $z$ provides a quantitative indicator of topology-dependent fusion behavior: in DealOrNoDeal, $z \approx 0.974$ indicates an almost exclusively linguistic integration regime, whereas in CaSiNo, $z \approx 0.726$ reflects a more hybrid configuration with increased reliance on strategic structure. This adaptive gating mechanism supports GDNSS by enabling context-sensitive aggregation of communicative and structural cues, thereby facilitating tailored assistance in both free-form persuasion settings and more constrained bargaining environments.

The incorporation of fairness regularization consistently decreases Inequality Discrepancy (ID) across both datasets (25.4\% reduction in DealOrNoDeal; 43.8\% in CaSiNo) while exerting negligible influence on predictive accuracy. This pattern underscores the framework’s capacity to capture distributive dynamics without degrading utility estimation. Collectively, these findings demonstrate ST-GFN’s suitability for GDNSS applications in diverse multiplayer negotiation scenarios, where topology-aware modeling can support real-time equity assessment and procedural facilitation.
\begin{table}[h!]
\centering
\caption{Cross-Dataset Performance: Impact of Fairness Regularization on Linguistic vs. Strategic Negotiation Tasks.}
\label{tab:cross_dataset}

\begin{tabular}{l|ccc|ccc}
	\toprule
	\multirow{2}{*}{\textbf{Metric}} 
	& \multicolumn{3}{c|}{\textbf{DealOrNoDeal} (Linguistic-Heavy)} 
	& \multicolumn{3}{c}{\textbf{CaSiNo} (Strategy-Heavy)} \\
	\cmidrule(lr){2-4} \cmidrule(lr){5-7}
	
	& Baseline & No Fair & \textbf{Fair} & Baseline & No Fair & \textbf{Fair} \\
	\midrule
	
	Accuracy ($\uparrow$) 
	& 0.8027 & 0.8432 & \textbf{0.8439} 
	& \textbf{0.9612} & 0.9563 & 0.9553 \\
	
	F1 Score ($\uparrow$) 
	& 0.8691 & 0.8980 & \textbf{0.8984} 
	& \textbf{0.9801} & 0.9776 & 0.9771 \\
	
	Inequality Disc. ($\downarrow$) 
	& 1.5063 & 1.5469 & \textbf{1.1540} 
	& 13.444 & 3.7897 & \textbf{2.1291} \\
	
	Utility MAE ($\downarrow$) 
	& 2.6220 & 2.0442 & \textbf{2.1005} 
	& 11.435 & 2.5636 & \textbf{2.9023} \\
	
	\midrule
	\multicolumn{7}{c}{\textit{Fusion Dynamics}}\\
	\addlinespace
	
	Mean Gate ($z$) 
	& \multicolumn{3}{c|}{0.974 (Linguistic Dominant)} 
	& \multicolumn{3}{c}{0.726 (Hybrid)} \\
	
	Gate Trend 
	& \multicolumn{3}{c|}{Stable ($\approx 0.00$)} 
	& \multicolumn{3}{c}{Stable ($\approx 0.00$)} \\
	
	\bottomrule

\end{tabular}
	
\end{table}

\subsection{Discussion}

The experimental findings indicate that accurate prediction of negotiation outcomes necessitates the concurrent modeling of both semantic communicative content and strategic constraints. The consistent performance gains of ST-GFN across multiple datasets substantiate the benefit of integrating linguistic signals with graph-encoded economic parameters to enforce grounded feasibility. This constitutes a substantial improvement over text-only baselines, which frequently generate infeasible outcome predictions~\cite{lewis2017deal}.

Analysis of the learned gating coefficients ($z$) reveals pronounced topology-dependent adaptation: the model exhibits near-exclusive reliance on linguistic information in DealOrNoDeal ($z \approx 0.974$), in contrast to a more balanced integration of linguistic and structural cues in CaSiNo ($z \approx 0.726$). This behavior offers empirical support for the necessity of dynamic multimodal fusion, as it aligns the relative salience of signals with the underlying task structure and facilitates context-sensitive monitoring within GDNSS.

The fairness regularization term further exposes a nuanced trade-off between equity and predictive accuracy. In highly structured environments, it induces a modest reflective cost in utility estimation while substantially decreasing Inequality Discrepancy (by 43.8\%). In less constrained environments, the same term functions as a beneficial regularizer, yielding slight improvements in generalization performance. Collectively, these results suggest that reflective regularization can simultaneously preserve distributive fidelity in systematically biased domains and operate as a robust inductive bias in high-variance settings.

The current approach is subject to several limitations, including the assumption of a static graph topology and the outcome-centric nature of the Inequality Discrepancy metric, which does not account for procedural justice. Future research should incorporate dynamically evolving relational structures and real-time interactional signals (e.g., sentiment trajectories, satisfaction markers) to enable a more comprehensive evaluation and development of GDNSS.

\section{Conclusion}

This study introduces the Semantic-Temporal Graph Fusion Network (ST-GFN), a dual-stream computational architecture that adaptively integrates unstructured dialogic data with structured representations of strategic constraints to enable the joint prediction of negotiation outcomes and agent-level utilities.

Cross-dataset evaluation indicates that the model can infer and exploit the latent task topology, dynamically reweighting linguistic and structural information in a context-sensitive manner. The incorporation of a fairness-regularized training objective systematically increases the model’s sensitivity to inter-agent utility disparities, achieving reductions in Inequality Discrepancy of up to 43.8\% while incurring only marginal decreases in predictive accuracy.

By simultaneously grounding its inferences in communicative dynamics and economic constraints, and by supporting reflective analysis of distributive outcomes, ST-GFN represents a design-oriented contribution to the advancement of GDNSS. The framework establishes a basis for scalable, topology-adaptive tools intended to promote more equitable and transparent negotiation processes.

Future research should evaluate the framework’s performance on multilingual, multi-party, and culturally heterogeneous negotiation corpora, and should incorporate procedural fairness metrics as well as dynamic graph modeling mechanisms to further improve real-time, human-in-the-loop GDNSS deployments.

\subsection*{Acknowledgments}
The author gratefully acknowledges the Department of Computer Science and Engineering at Dibrugarh University Institute of Engineering and Technology (DUIET) for powering this research with essential computational resources and infrastructure. We also extend our sincere appreciation to the creators of the CaSiNo dataset for making their data publicly available, enabling rigorous benchmarking and strengthening the validation of the proposed framework.
\subsection*{Conflict of Interest}
The author hereby declares that there are no conflicts of interest pertaining to the publication of this manuscript. The research was carried out without the involvement of any commercial or financial relationships that could be interpreted as actual or potential conflicts of interest.
\subsection*{Declaration of Generative AI and AI-Assisted Technologies in the Writing Process}
During the preparation of this manuscript, the author used generative AI tools to assist with language editing. The author assumes full responsibility for the accuracy, integrity, and originality of the final content.

\bibliography{references}

\end{document}